\documentclass[pre,floatfix,reprint,longbibliography,numerical]{revtex4-1}%
\usepackage[utf8]{inputenc}
\usepackage[english]{babel}
\usepackage[T1]{fontenc}
\usepackage{amssymb,amsmath,amsthm}
\usepackage{graphicx}
\usepackage{bm}
\usepackage{color}
\usepackage[normalem]{ulem}
\definecolor{darkblue}{rgb}{0,0,.5}
\usepackage[linktocpage, colorlinks=true,linkcolor=darkblue, citecolor=darkblue]{hyperref}

\newcommand{\de}{\partial}

\newcommand{\vc}[1]{\bm{#1}}
\newcommand{\vt}[1]{\bm{\mathsf{#1}}}
\newcommand{\tsp}{\mathsf{T}}

\newcommand{\tr}{\operatorname{\mathrm{tr}}}

\newcommand{\ed}{\dot{\varepsilon}}
\newcommand{\dv}{\hat{\vc d}}

\newcommand{\tensprod}{{}}

\newcommand{\cauchy}{\vt{T}}

\begin{document}

\title[Rheometric framework for generic steady flows]{A theoretical framework for steady-state rheometry in generic flow conditions}

\author{Giulio G.~\surname{Giusteri}}
\email{Current address: Dipartimento di Matematica, Politecnico di Milano, Piazza Leonardo da Vinci 32, 20133 Milano, Italy. E-mail: giulio.giusteri@gmail.com}
\affiliation{Mathematics, Mechanics, and Materials Unit, Okinawa Institute of Science and Technology Graduate University, \mbox{1919-1 Tancha, Onna, Okinawa, 904-0495, Japan}}
\author{Ryohei Seto}
\email{setoryohei@me.com}
\affiliation{Mathematics, Mechanics, and Materials Unit, Okinawa Institute of Science and Technology Graduate University, \mbox{1919-1 Tancha, Onna, Okinawa, 904-0495, Japan}}

\date{\today}

\begin{abstract}
We introduce a general decomposition of the stress tensor for incompressible fluids in terms of its components on a tensorial basis adapted to the local flow conditions, which include extensional flows, simple shear flows, and any type of mixed flows.
Such a basis is determined solely by the symmetric part of the velocity gradient and allows for a straightforward interpretation of the non-Newtonian response in any local flow conditions.
In steady homogeneous flows, the material functions that represent the components of the stress on the adapted basis generalize and complete the classical set of viscometric functions used to characterize the response in simple shear flows.
Such a general decomposition of the stress is effective in coherently organizing and interpreting rheological data from laboratory measurements and computational studies in non-viscometric steady flows of great importance for practical applications.
The decomposition of the stress in terms with clearly distinct roles is also useful in developing constitutive models.
\end{abstract}

\maketitle

\section{Introduction}

The typical workflow associated with the mathematical modeling of physical phenomena starts with the collection of experimental data.
These can originate from laboratory measurements or from computational studies based on ``lower level'' physics, for which reliable and well-tested models are already available.
Given the data, two immediate challenges concern their organization and interpretation.
Most often, the interpretation of experimental data involves making a connection with the specific conditions under which phenomena were observed, meanwhile tracing the limits of validity of the conclusions that can be drawn.

The study of rheological properties of fluids and their mathematical modeling follow this general scheme, with an emphasis on the main next step: The interpolation and extrapolation of the collected data.
In fact, there is a strong technological interest in controlling the behavior of fluids in a variety of flow regimes that are not easily accessible to experimental measurements.
The extrapolation of collected data to such regimes is at the heart of constitutive modeling, where physical insight and mathematical tools come together with the ultimate goal of providing reliable simulations of engineering-scale flows of complex fluids.

Rheological measurements are challenging and the identification of suitably controllable flows is a crucial issue.
In this respect, of great importance is the class of viscometric flows, that provided the basic platform also for the interpretation of measurements.
Within that framework, three material functions---shear viscosity $\eta_\mathrm{S}$ and normal stress differences $N_1$ and $N_2$---are shown to characterize the fluid response in viscometric flows as the shear rate is varied \cite{Coleman_1966,Macosko_1994,Larson_1999}.
Nevertheless, their applicability is limited and relies on the identification of locally co-moving frames in which the velocity gradient resembles that of a simple shear.

\begin{figure}[b]
\includegraphics[width=0.5\textwidth]{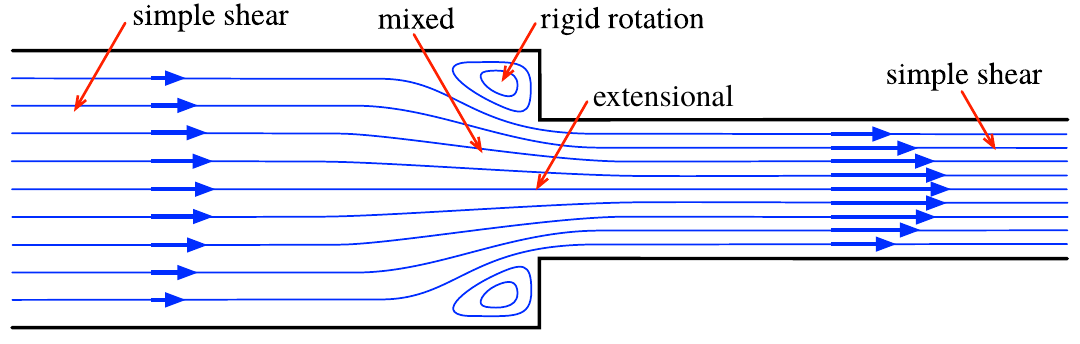}
\caption{The channel flow through a contraction offers a paradigmatic example of non-viscometric flow. The gradient of the velocity field in such a steady flow is equivalent to that of extensional flows at the centerline of the contracting region, of shear flows far from the contracting region, of a rigid rotation at the center of the vortices, and of mixed flows in the intermediate regions.}\label{fig:contraction}
\end{figure}

However, non-viscometric flows are usually observed in most physical systems, hence the need for exploring the fluid response in flow conditions other than viscometric ones.
A paradigmatic example of non-viscometric flow is the channel flow through a contraction (Fig.~\ref{fig:contraction}). 
The sudden reduction of the channel width leads to an increase in the streaming velocity at the center of the channel and to the appearance of counterrotating vortices in the corners right before the contraction.
In contrast to what happens in a channel with uniform width, where the velocity gradient is everywhere equivalent to that of a simple shear, the gradient of the velocity field in a steady flow through a contraction is equivalent to that of extensional flows at the centerline of the contracting region, of shear flows far from the contracting region, of a rigid rotation at the center of the vortices, and of mixed flows in the intermediate regions.

The widespread occurrence of similar flow conditions in real systems has prompted the study of rheological properties of fluids especially in extensional flows \cite{McKinley_2002,Petrie_2006,Dai_2017} and the development of computational techniques able to access extensional flows and mixed flows \cite{Kraynik_1992,Todd_1998,Baranyai_1999,Hunt_2010,Dobson_2014,Zinchenko_2015,Jain_2015,Hunt_2016,Cheal_2018}, which range from pure extension, to simple shear, to rigid rotation of the fluid.
Through similar studies it has long since become clear that the non-Newtonian fluid response often depends on the local flow type (extensional, simple shear, or mixed). 

The need for dealing with controllable flows when exploring non-Newtonian responses makes it desirable to generate uniform flow conditions, but this is not always possible. 
As for the planar case, it is known that the four-roll mill apparatus devised by Taylor \cite{taylor_1934} can be used to produce any of the mixed flows in a significant neighborhood of the stagnation point \cite{Fuller_1981,Larson_1999}.
This provided clear hints for the design of cross channels that produce the same type of flows \cite{Hudson_2004,Lee_2007,Lee_2007a,Deschamps_2009,Haward_2012}.
Nevertheless, not all of the interesting flow conditions can be investigated through flows with uniform velocity gradient.
This stimulated the development and application of rheo-optical techniques \cite{Fuller_1995,Pathak_2006,Haward_2012,Ober_2013,Shribak_2015,Zhao_2016,Sun_2016}, designed to provide local measurements of the stress generated in possibly non-uniform flows.
Techniques that provide local velocity measurements under controlled stress conditions \cite{Callaghan_1999,Besseling_2007,Manneville_2008,Dimitriou_2012,Gallot_2013,Saint-Michel_2016} contribute in an important complementary way to the understanding of the local response of complex fluids.

In spite of the vast amount of data now available in non-viscometric flow conditions, the lack of a general scheme to organize and interpret such data has led to the introduction of various quantities that are connected via \emph{ad hoc} relations to the classical viscometric functions (viscosity and normal stress differences) associated with simple shear flows. 
The main aim of the present paper is to provide a new scheme for the organization and interpretation of rheological measurements for steady flows of incompressible fluids.
In particular, we show how to define in a unified way the material functions that are needed to describe the local fluid response in non-viscometric flows such as the contraction flow of Fig.~\ref{fig:contraction}.
This is achieved by introducing, in Sec.~\ref{sec:representation}, general response coefficients, each representing a distinct degree of freedom of the Cauchy stress tensor.

Subsequently,  in Sec.~\ref{sec:flow-class}, a complete set of material functions is associated with the response coefficients. 
Our scheme goes beyond the one given by viscometric functions and it is complete in the sense that it gives a coherent interpretation of data obtained for any three-dimensional flow.
This rheometric framework has been applied by the authors in a recent computational study of dense suspensions \cite{Seto_2017}.

Finally, in Sec.~\ref{sec:constitutive}, we discuss how the material functions can be used in developing constitutive models. 
This is done by reinterpreting well-known models in terms of those functions and suggesting further ways to exploit the physical insight associated with the general decomposition of the stress tensor provided in Sec.~\ref{sec:representation}.

\section{Local decomposition of the stress tensor}\label{sec:representation}

Our main result is the construction and interpretation of a general decomposition of the stress tensor, given in Eq.~\eqref{eq:T}, for an incompressible fluid motion. 
Such a decomposition associates the six degrees of freedom of the symmetric Cauchy stress with distinct effects. 
This is achieved by projecting the stress, at each point in space and instant in time, on a tensorial basis which is adapted to the local description of the flow in terms of the symmetric part $\vt D$ of the velocity gradient
\begin{equation}\label{eq:decomp}
\nabla\vc u=\vt D+\vt W,
\end{equation}
with $\vt W$ denoting its antisymmetric part.

{The choice of constructing the tensorial basis starting from $\vt D$ is most effective when it is physically informative to organize the degrees of freedom of the stress in relation to those of $\vt D$.
This is particularly true whenever the stress is chiefly determined by how the material is flowing, since $\vt D$ encodes essential information concerning the deformation rate associated with the flow.
As we discuss in Sec.~\ref{sec:standard}, this choice enables us to generalize and complete the standard definition of material functions for steady homogeneous flows.
Nevertheless, the general decomposition~\eqref{eq:T} introduced below is applicable and provides useful information even in unsteady flows.
We discuss in Sec.~\ref{sec:nonD} situations in which different starting points for the stress decomposition may be helpful.}

\subsection{Parametrization of the velocity gradient}

We consider situations in which the symmetric tensor $\vt D$ has a nonvanishing dominant eigenvalue (with largest absolute value).
%
We further denote by $\dv_1$ the unit-norm eigenvector of $\vt D$ corresponding to the dominant eigenvalue.
In the particular cases in which $\vt D$ has two dominant eigenvalues, corresponding to the planar flows discussed below, we fix $\dv_1$ by choosing $\ed$ in Eq.\,\eqref{eq:D-3d} 
as the positive eigenvalue.
(However, $\ed$ can be negative for generic three-dimensional flows.)
Using the eigenvalues and eigenvectors of $\vt D$ and from the angular frequency and the axis of the rigid rotation associated with $\vt W$ (quantities that are defined without reference to any choice of coordinate system), it is possible to represent the eight degrees of freedom that characterize the velocity gradient in any incompressible fluid motion.

Using the decomposition \eqref{eq:decomp} and writing $\vc a\vc b$ for the dyadic product of vectors $\vc a$ and $\vc b$, the most general traceless velocity gradient is given by 
\begin{equation}\label{eq:D-3d}
\vt D = 
\frac{2\ed}{\sqrt{3+4\alpha^2}}
\bigl[
\dv_1\tensprod\dv_1-(1/2+\alpha)\dv_2\tensprod\dv_2-(1/2-\alpha)\dv_3\tensprod\dv_3
\bigr],
\end{equation}
where the dimensionless asymmetry parameter $\alpha$ ranges from $0$ to $1/2$, and $\dv_1$, $\dv_2$, and $\dv_3$ are orthonormal eigenvectors of $\vt D$, and by
\begin{equation}
\vt W=
\ed
\Bigl[
  \beta_1\bigl(\dv_3\tensprod\dv_2-\dv_2\tensprod\dv_3\bigr) 
+ \beta_2\bigl(\dv_1\tensprod\dv_3-\dv_3\tensprod\dv_1\bigr)
+ \beta_3\bigl(\dv_2\tensprod\dv_1-\dv_1\tensprod\dv_2\bigr)\Bigr],\label{eq:W-3d}
\end{equation}
with $\beta_k$ being a dimensionless parameter for each $k=1,2$, and $3$. These encode the angular frequency and the axis of the rotation associated with $\vt W$. Indeed, $2\ed\beta_k$ corresponds to the component along $\dv_k$ of the vorticity vector, namely
\begin{equation}
\beta_k=\frac{1}{2\ed}\dv_k\cdot\nabla\times\vc u.
\end{equation}
For any value of the dimensionless parameters $\alpha$ and $\beta_k$ ($k=1,2$, and $3$), 
the local timescale of the deformation is set by the rate 
$ |\ed| = \sqrt{\tr(\vt D^2)/2}$. 

Without loss of generality, we have associated the eigenvector $\dv_3$ with the eigenvalue with least absolute value.
This choice has the advantage that the velocity in planar flows has components only in the plane spanned by $\dv_1$ and $\dv_2$.

\subsection{Adapted tensorial basis}

Our objective is to construct a decomposition of the stress tensor which is adapted to the local flow. To this end, we define an orthogonal basis for symmetric tensors built starting from the identity tensor $\vt I$ and $\vt D$. We remark that such a basis is completely independent of $\vt W$.

Since $\vt D$ is traceless (due to the incompressibility constraint) then it is orthogonal to $\vt I$. The subspace of symmetric tensors that are diagonal on the basis of the eigenvectors of $\vt D$ is three-dimensional. We then need to find only one tensor $\vt E$ which is orthogonal to $\vt I$ and $\vt D$ and diagonal on the basis $(\dv_1,\dv_2,\dv_3)$. This can be easily shown to be
\begin{equation}
\vt E=
\frac{\ed}{\sqrt{3+4\alpha^2}}
\bigl[
-2\alpha
\dv_1\tensprod\dv_1
- (3/2 - \alpha)
\dv_2\tensprod\dv_2
+ (3/2 + \alpha)
\dv_3\tensprod\dv_3\bigr],
\end{equation}
where we have chosen the normalization factor in such a way that
$\sqrt{\tr(\vt E^2)}= \sqrt{3/2}|\ed|$.
%
%
To complete the basis, we can simply consider the three off-diagonal tensors
\begin{equation}
\vt G_i=\ed\bigl(\dv_j\dv_k+\dv_k\dv_j\bigr),
\end{equation}
with $i\neq j\neq k$ ranging from $1$ to $3$.

We now introduce dimensionless tensor fields
\begin{equation}\label{eq:hattensors}
\hat{\vt D}=\ed^{-1}\vt D,\quad\hat{\vt E}=\ed^{-1}\vt E,\quad\hat{\vt G}_i=\ed^{-1}\vt G_i,
\end{equation}
so that we can identify a dimensionless adapted basis
\begin{equation}\label{eq:basis}
\mathcal B=\big(\vt I,\hat{\vt D},\hat{\vt E},\hat{\vt G}_1,\hat{\vt G}_2,\hat{\vt G}_3\big).
\end{equation}
All of these tensors are orthogonal to each other in the sense that $\tr(\vt A^\tsp\cdot\vt B)=\vt A:\vt B=0$ for any choice of $\vt A$ and $\vt B$ in $\mathcal B$. Such a basis depends locally on space and time through the eigenvalues and eigenvectors of $\vt D$ and is not defined whenever $\vt D$ vanishes.

\subsection{Response coefficients}

\begin{figure*}
\includegraphics[width=0.98\textwidth]{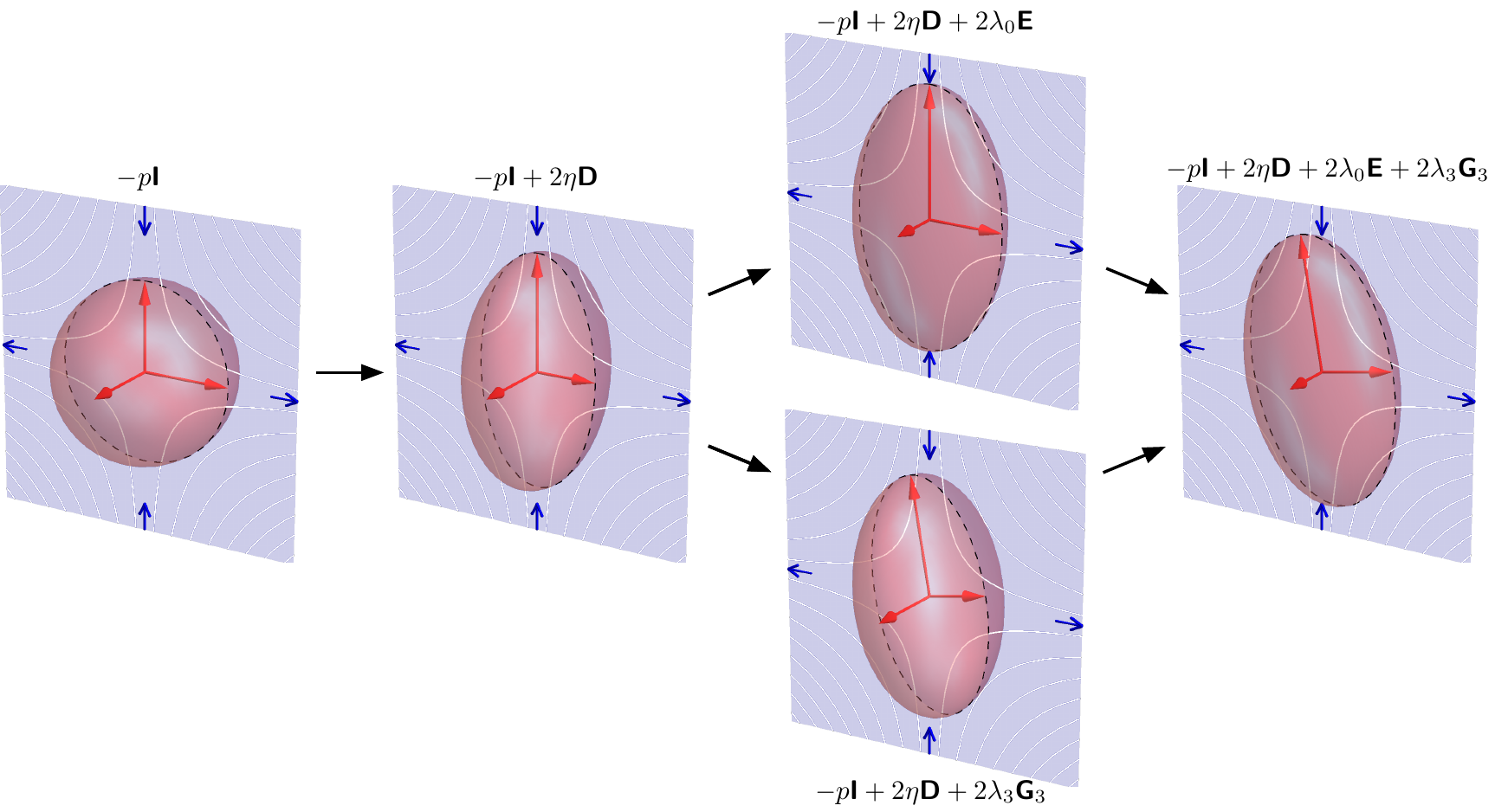}
\caption{Each of the elements of the adapted tensorial basis $\mathcal B$ introduced 
in Eq.\,\eqref{eq:basis} encodes a different degree of freedom of the stress tensor $\cauchy$.
The response coefficients $p$, $\eta$, $\lambda_0$, $\lambda_1$, $\lambda_2$, and $\lambda_3$ measure the relevance of each degree of freedom.
Considering planar flows, $\eta$ describes the anisotropy of the stress tensor in the flow plane and it is associated with the rate at which mechanical energy is being converted into internal energy.
The response coefficient $\lambda_0$ governs the out-of-flow-plane anisotropy, modifying the eigenvalues but keeping the eigenvectors of the stress aligned to those of $\vt D$.
The response coefficients $\lambda_1$, $\lambda_2$, and $\lambda_3$ generate a rotation of the eigenvectors of the stress with respect to those of $\vt D$.
The length of the axes of the ellipsoids in each panel represents the absolute value of the stress eigenvalues, 
while the arrows within the ellipsoids show the direction of the eigenvectors.}\label{fig:material-functions}
\end{figure*}

The stress tensor $\cauchy$ can be decomposed on the basis $\mathcal B$ as
 \begin{equation}\label{eq:T}
\cauchy=-p\vt I+2\ed\bigl(\eta\hat{\vt D}
+\lambda_0\hat{\vt E}
+\lambda_1\hat{\vt G}_1+\lambda_2\hat{\vt G}_2+\lambda_3\hat{\vt G}_3\bigr).
\end{equation}
Each of the response coefficients $p$, $\eta$, $\lambda_0$, $\lambda_1$, $\lambda_2$, and $\lambda_3$ is affected, in principle, by the value of any quantity that describes the state of the system.
Indeed, from Eq.~\eqref{eq:T} and the definitions \eqref{eq:hattensors} we can easily infer that 
\begin{equation}
p=-\frac{\cauchy:\vt I}{\vt I:\vt I}=-\frac{1}{3}\tr(\cauchy)
\end{equation}
and 
\begin{equation}\label{eq:rf}
  \eta=\frac{1}{2}\frac{\cauchy:\vt D}{\vt D:\vt D},\quad
  \lambda_0=\frac{1}{2}\frac{\cauchy:\vt E}{\vt E:\vt E},\quad\text{and}\quad
  \lambda_k=\frac{1}{2}\frac{\cauchy:\vt G_k}{\vt G_k:\vt G_k}
\end{equation}
for $k=1,2$, and $3$, showing that, while the basis $\mathcal B$ depends only on $\vt D$, the response coefficients are influenced by anything that affects the stress tensor $\cauchy$. {We used the symbol $\eta$, commonly employed to denote the shear viscosity, for one of our material coefficients, since the latter is a generalization of the shear viscosity and, in a steady simple shear, it becomes \emph{exactly} the shear viscosity. Nevertheless, it is important to observe that the coefficient $\eta$ is defined in a broader sense than the shear viscosity, being meaningful in many different flow conditions.
We also note that, while $\vt\sigma$ is a more common notation for the stress tensor, we prefer to use $\cauchy$ and consistently employ greek letters for scalars, lower-case bold for vectors, and upper-case bold for tensors.}

The major advantage of decomposing the stress $\cauchy$ according to Eq.~\eqref{eq:T} is that each of the response coefficients has a precise role that is independent of any specific flow conditions.
The coefficient $p$ clearly measures the isotropic pressure contribution to the stress. 
The scalar product $\cauchy:\vt D$ that defines $\eta$ measures the rate at which mechanical energy is being converted into internal energy.
In view of this, $\eta$ contains not only the contribution due to the irreversible dissipation of mechanical energy (normally associated with viscosity) but also the reversible storage of kinetic energy into internal elastic energy \cite[Sec.~6]{Pasquali_2004}.
We can thus consider $\eta$ as a generalized viscosity.
The response coefficients $\lambda_k$ ($k=0,\ldots,3$) have a simple interpretation based on how the eigenvectors and eigenvalues of the stress tensor $\cauchy$ are related to those of $\vt D$ (Fig.~\ref{fig:material-functions}).
If they all vanish identically (but $\eta$ is nonzero) $\cauchy$ has the same eigenvectors of $\vt D$ and proportional eigenvalues.
If only $\eta$ and $\lambda_0$ are nonvanishing, the eigenvectors of $\cauchy$ are still aligned with those of $\vt D$, but the relative magnitude of the corresponding eigenvalues is no longer the same.
Hence, the presence of $\lambda_0$ means that the intensity of the stress is not distributed along its principal directions proportionally to the distribution of the rate of deformation.
In the presence of nonvanishing $\lambda_1$, $\lambda_2$, or $\lambda_3$, the eigenvectors of $\cauchy$ are no longer aligned with the eigenvectors of $\vt D$, a phenomenon typically associated with elastic effects but also with modifications in the microstructure of complex fluids.

The above construction and arguments are all local in nature, meaning that each quantity can take different values at different points in space and instants in time.
Nevertheless, the meaning of the response coefficients is everywhere the same.
This is a key feature of the decomposition \eqref{eq:T} in comparison to other possible ways of representing the six degrees of freedom of the symmetric stress tensor $\cauchy$ (such as, for instance, listing its components on a fixed orthonormal basis using the lab frame).

Organizing the data collected (experimentally or computationally) about the local stress by means of the response coefficients introduced above is particularly useful when one needs to compare the fluid behavior observed in different conditions and geometries, because the data structure is built with reference to the same physical facts.
The use of the tensor $\vt D$ as starting point for the definition of a tensorial basis is very convenient, since it is often possible to have fairly accurate local measurements of the velocity gradient.
Nevertheless, this does not entail any restriction about the nature of the independent quantities that affect the stress tensor. 
{The latter can depend, for instance, on the history of the strain as well as the strain rate, as needed in the presence of elastic effects.}

\subsection{{Remarks on the treatment of elastic effects}}\label{sec:nonD}

Even though the decomposition \eqref{eq:T} is applicable and can give useful information also in the presence of elastic effects, in many rheological experiments that explore such elastic properties one can encounter situations in which $\vt D$ vanishes and the tensorial basis \eqref{eq:basis} is not naturally defined.
Nevertheless, this issue can be easily overcome in a number of situations based on the following argument.
If $\vt D=\vt 0$ in a finite three-dimensional region, then the material there is moving rigidly, or is not moving at all.
This means that, in a generic flow, the tensor $\vt D$ can vanish only at isolated points, along lines, or on some surface---typical examples are the axes of vortices where the flow tends to a rigid rotation.
In such cases it is usually possible to extend the definition of the dimensionless tensorial basis \eqref{eq:basis} to the region where $\vt D=\vt 0$ by continuity with that in the neighboring points.

Second, there are important oscillatory flows in which the velocity uniformly vanishes at the periodic turning points of the flow.
Also in this case, since the static condition appears only at isolated instants, it is usually easy to extend by continuity the definition of the tensorial basis \eqref{eq:basis} to those instants.
Considering, for example, oscillatory shear flows, the normalized basis tensors are constant in time and homogeneous in space, always well-defined except at the turning instants, but trivially extendible even there.
This shows how the decomposition \eqref{eq:T} maintains its efficacy beyond the context of steady homogenous flows.

Finally, there are experiments, used to study stress relaxation phenomena, in which a shearing flow is stopped and then $\vt D$ vanishes.
In this cases, it is clearly necessary to decompose the stress on bases determined by tensorial quantities other than $\vt D$.
Measures of an effective stretching, such as the Finger tensor or the commonly used conformation tensor, would be an appropriate starting point. 
We do not exclude the possibility that some of the tensorial bases developed in a similar context may provide useful insight even in more dynamical situations. 
An interesting contribution in this direction can be found in a paper 
of Pasquali and Scriven \cite{Pasquali_2002}, where $\vt D$ is projected onto the eigenvectors of the conformation tensor (associated with the elastic stress) to define a ``molecular extension rate'' and a ``molecular shear rate''.

\section{Flow classifications and constitutive modeling}\label{sec:flow-class}

The primary purpose of the stress decomposition \eqref{eq:T} is a systematic organization of rheological measurements in different local flow conditions.
Nevertheless, it can also be useful in developing constitutive models.
Here we do not discuss new constitutive models, but we want to highlight the contexts 
in which Eq.~\eqref{eq:T} is most useful and its compatibility with any existing constitutive model.

The main steps in this respect involve, first, a discussion of how the choice of independent descriptors affects the local flow classification and, second, upgrading the response coefficients introduced above to material functions, in terms of which constitutive models can be formulated.
In this section, we also clarify how the new material functions are a generalization of standard material functions, such as viscometric functions and extensional viscosity.

\subsection{Independent descriptors and flow types}

The choice of the independent fields, the values of which characterize the local state of a system, is clearly the first step in constitutive modeling.
There is no \emph{a priori} indication that one can give about this choice, except that one would like to use quantities that are measurable and to avoid redundancy, in the sense that two distinct sets of values should label distinct kinematical states of the system.

In a rheological context, once this initial choice is performed we can also ask whether distinct (local) kinematical conditions could or should be considered equivalent in view of the fact that one may expect to measure the same stress in different conditions.
This marks, in our opinion, a useful distinction between mere parametrizations of the local flow conditions (in terms of the chosen independent fields) and local flow-type classifications, which separate the local flow conditions in various equivalence classes based on reasonable expectations about the material response.  

Flow classification criteria have a long history (nicely reviewed 
by Thompson and Souza Mendes~\cite{Thompson_2005}) but the distinction between kinematic parametrization and flow classification remains often implicit.
It is nevertheless important to realize that each flow classification scheme incorporates some constitutive choices and expectations in the way local kinematic conditions are considered equivalent or not.

A seminal contribution concerning flow classifications was given by Astarita~\cite{Astarita_1979}, who lists locality, applicability to generic flows, and objectivity as important properties of flow classifications.
He motivates the requirement of objectivity (namely, covariance under possibly time-dependent changes of observer corresponding to the group of rigid-body motions) by saying that \emph{``Since the main reason for classifying flow fields is to decide which constitutive equation is more likely to produce useful results, the criterion should enjoy the same invariance properties that are required of the constitutive equation''}.
We share Astarita's view that a flow classification is intimately linked to some constitutive properties, but this opens the possibility of encountering situations in which the flow classification criterion should not be objective, but only Galilean covariant.

Indeed, the main point of a classification is to group conditions that are, at the microscopic level, physically equivalent and entail equivalent material responses at the macroscopic level.
In view of this, two microscopic situations that are mapped onto each other by means of \emph{accelerating} changes of observer, that are included in the objectivity requirement, can be equivalent only if inertial effects are negligible at the microscopic level (as discussed by Beris and Edwards~\cite[Sec.~7.2.1]{Beris_1994} and Phan-Thien~\cite[Sec.~4.3]{Phan-Thien_2002}).
This is generally the case for the type of elastic fluids that motivated Astarita's analysis and for many non-Newtonian fluids, but suspensions of (density-mismatched) particles in viscous fluids are the simplest examples of systems in which microinertial effects may be relevant and the constraint of objectivity too strong \cite{Ryskin_1980}. 

\begin{figure*}
\includegraphics[width=1.\textwidth]{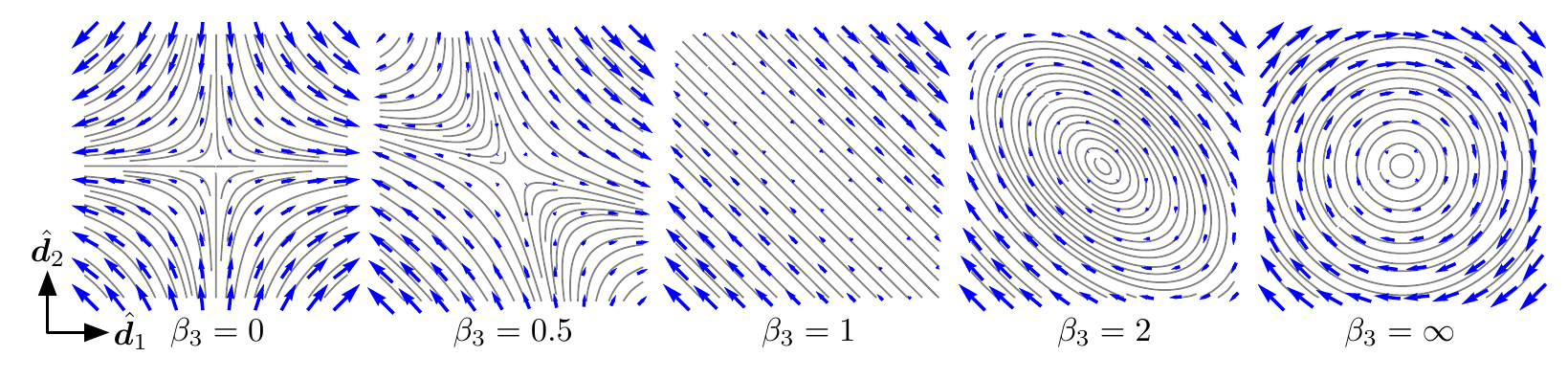}
\caption{{In homogeneous planar flows, the flow type is controlled by the dimensionless parameter $\beta_3$ which measures the relative importance of vorticity.}
Through mixed flows, we can interpolate between purely extensional flows ($\beta_3=0$), simple shear ($\beta_3=1$), and rigid rotation ($\beta_3=\infty$). In the latter case, a limit in which $\ed\to0$, so that the rotation rate $\ed\beta_3$ remains finite, is understood.
The extension and contraction axes are always identified by the eigenvectors $\dv_1$ and $\dv_2$ of $\vt D$.
}\label{fig:flow-types}
\end{figure*}

Here we give an example of how different flow classifications can be applied in similar situations in relation to different constitutive assumptions.
We begin by considering the class of steady homogenous planar flows, which can be \emph{parametrized} simply by the structure of the velocity gradient.
With reference to the representations \eqref{eq:D-3d} and \eqref{eq:W-3d} of $\vt D$ and $\vt W$, such flows correspond to choosing a fixed $\ed$, $\alpha=1/2$ (maximal asymmetry) and $\beta_1=\beta_2=0$ with $\beta_3$ constant (vorticity orthogonal to the flow plane).
Since there are two dominant eigenvalues of $\vt D$, we select $\dv_1$ so that $\ed>0$, as mentioned above.
Within this class of flows, simple shear and (planar) extensional flow emerge by choosing $\beta_3=1$ and $\beta_3=0$, respectively, while for any other value of $\beta_3$ the flow is mixed.
Notably, a rigid rotation of the fluid is approached for $\beta_3\gg 1$ and the streamlines are elliptical for $\beta_3>1$ (Fig.~\ref{fig:flow-types}).

It is possible to construct two purely kinematic flow classifications (one objective and the other one non-objective) in which the homogeneous planar flows remain distinct but the local flow conditions in inhomogeneous flows are grouped in different ways as regards their local equivalence to homogeneous flows. 
The fact that a classification is able to distinguish among different homogeneous planar flows is very important for all those fluids that display flow-type dependence by behaving differently in extensional and simple shear flows.

A local flow classification able to distinguish between extensional and simple shear flow and based \emph{exclusively} on the components of the velocity gradient can be established, by using the local flow parametrization given above, in terms of the rate $\ed$ and the component $\beta_3$ of the vorticity.
This classification is not objective, since any value of $\beta_3$ can be mapped to zero by considering spinning observers.
Hence, this classification might be useful when microinertial effects are relevant, but it is not otherwise appropriate.

To obtain an objective local flow classification able to distinguish between extensional and simple shear flow it is necessary to go beyond the exclusive use of the velocity gradient.
We recount here the presentation given by Schunk and Scriven \cite{Schunk_1990} of a quite simple but effective scheme. 
The set of kinematical parameters used for the classification is expanded to include (some of) the degrees of freedom associated with the rate of change of the tensor $\vt D$ along streamlines.
In particular, the local spin of the eigenvectors of $\vt D$ is encoded in the vector
\begin{equation}\label{eq:Dspin}
\vc w = \frac{1}{2}\sum_k\dv_k\times\bigg(\frac{\de\dv_k}{\de t}+\vc u\cdot\nabla\dv_k\bigg).
\end{equation}
The new degrees of freedom are the components of $\vc w$ on the eigenvectors of $\vt D$, suitably normalized.
We thus introduce, for $k=1,2$, and $3$, the dimensionless parameters
\begin{equation}\label{eq:dspin}
\delta_k=\frac{1}{\ed}\dv_k\cdot\vc w.
\end{equation}
An objective flow-type parameter for planar flows can be identified with the relative rotation parameter $\bar\beta_3=\beta_3-\delta_3$.
In this way, planar flows can be locally classified by using the rate $\ed$ and $\bar\beta_3$.

In steady homogeneous flows, the eigenvectors of $\vt D$ do not rotate and $\delta_3$ vanishes, so that $\bar\beta_3=\beta_3$.
As a consequence, homogeneous planar flows are equally distinguished by the non-objective scheme and the objective one.
A striking difference, however, arises when considering non-homogeneous flows.

As a paradigmatic example, we can analyze the Newtonian profile for the tangential velocity component $u$ of a steady viscometric flow between concentric cylinders.
The objective classification assigns $\bar\beta_3=1$ everywhere in the flow domain (while $\ed$ is constant along streamlines), showing that, in the absence of microinertial effects, the flow is locally equivalent to simple shear.
On the other hand, the non-objective classification scheme gives a local flow-type that depends on the radial coordinate $r$ according to
\begin{gather}
  \ed(r) = \frac{1}{2}\bigg(\frac{\de u}{\de r}(r)-\frac{u(r)}{r}\bigg),\\
  \beta_3(r) = \frac{r\frac{\de u}{\de r}(r)+u(r)}{r\frac{\de u}{\de r}(r)-u(r)},
\end{gather}
which is not equivalent to a simple shear, except at those points in which $u$ vanishes, if any.
This indicates how flows between rotating cylinders can provide interesting tests to assess the presence of microinertial effects, based on analyzing the compatibility of their behavior with the two different classification schemes.

\subsection{From response coefficients to material functions}

Once a choice of independent descriptors and local flow classification, appropriate to a specific class of fluids, is made, it is possible to associate each of the response coefficients introduced in Sec.~\ref{sec:representation} with a distinct material function.
The local flow classification labels distinct experimental conditions, so we can build material functions by interpolating the values of the response coefficients measured by varying the experimental conditions.

Nowadays, computational experiments offer important insight about the material response in conditions that are not easy to handle in a laboratory experiment.
Nevertheless, the range of local flow conditions that are encountered in real flow geometries remains wider than that accessible by well-controlled experiments.
This makes it necessary the extrapolation of the measured behavior to those conditions, which is at the heart of constitutive modeling, and material functions play an essential role in this.

The material functions $p$, $\eta$, and $\lambda_k$ (for $k=0,\ldots,3$) have the remarkable properties of identifying the same degrees of freedom of the stress in any local flow condition and of being applicable in conjunction with any flow classification scheme by simply changing the independent fields on which they depend.
The first property is particularly desirable due to the important increase of interest and available data in flows other than viscometric ones.
Standard material functions are indeed defined with reference to a specific flow type (for instance, simple shear or uniaxial and biaxial extension) with different choices for the reference deformation rate and the basis on which the stress tensor is decomposed.
Our material functions overcome these issues, since they are linked to projections of the stress along the eigenvectors of $\vt D$ (coordinate-independent and objective quantities) and are normalized by the rate $\ed$, which is defined in the same way for any local flow type. 

The second property highlights the general applicability of such material functions, because it shows that all the modeling assumptions come after their definition, which is based on the decomposition \eqref{eq:T} of the stress tensor.
This is also a property of standard viscometric functions and it is the basis of their enormous practical importance in rheology.
As a matter of fact, our material functions include and generalize the standard material functions used to characterize steady flows, by providing a unified tool to compare data across different flow conditions and different classes of fluids.

\subsection{Relation to standard material functions}\label{sec:standard}

To provide a clear connection with standard treatments and to highlight once more what type of generalization we have introduced, we show how viscometric functions and extensional viscosity are recovered within the new framework.
Since viscometric functions are defined for the planar simple shear flow, we also confine attention to the planar extensional flow, but the argument extends easily to three-dimensional extensional flows.
Moreover, since the standard material functions are usually employed to discuss fluid models in the absence of microinertial effects, we use the simple objective flow classification given above in which simple shear corresponds to the value $\bar\beta_3=1$ of the relative rotation parameter, while planar extension corresponds to $\bar\beta_3=0$.
Then, in this particular case, our material functions depend on the local rate $\ed$ and the flow-type parameter $\bar\beta_3$. 

If all of the physical effects that describe a system are invariant under translations in the direction orthogonal to the flow plane we can reasonably assume the planarity of the flow. 
In this case, a significant reorientation of the stress eigenvectors with respect to the eigenvectors of $\vt D$ can only take place in the flow plane, entailing $\lambda_1=\lambda_2=0$.
The general decomposition \eqref{eq:T} becomes now the representation
\begin{multline}
\cauchy(\ed,\bar\beta_3)  = 
-p(\ed,\bar\beta_3)\vt I  \\
+ 2\ed\bigl[\eta(\ed,\bar\beta_3)\hat{\vt D}+\lambda_0(\ed,\bar\beta_3)\hat{\vt E}+\lambda_3(\ed,\bar\beta_3)\hat{\vt G}_3\bigr].\label{eq:S-2d}
\end{multline}
{The basis tensors $\vt I$, $\hat{\vt D}$, $\hat{\vt E}$, and $\hat{\vt G}_3$, being defined in terms of the eigenvectors of $\vt D$, are all objective quantities.
Hence, the representation~\eqref{eq:S-2d} is objective if and only if the material functions $p$, $\eta$, $\lambda_0$, and $\lambda_3$ are objective.}

We stress that the local vanishing of $\bar\beta_1$ and $\bar\beta_2$ (always true in planar flows) does not necessarily cause $\lambda_1$ and $\lambda_2$ to vanish.
Rather, the presence of a stress associated with those quantities would render the planar flow conditions unstable, leading to more complex dynamics.
Similarly, in a stable extensional flow ($\bar\beta_3=0$) we would expect to find $\lambda_3=0$, but the presence of a nonvanishing $\lambda_3$ (possibly generated by elastic effects) could break the symmetry and destabilize the flow, as experiments suggest \cite{Haward_2016}.

The pressure $p$ retains its usual role of measuring the isotropic stress. 
For incompressible fluids, it combines the result of various microscopic effects with a reaction (or Lagrange multiplier) associated with the incompressibility constraint.
For this reason, it cannot be fully given by constitutive prescriptions.
In steady viscometric or homogeneous flows, since the internal energy of each fluid parcel is constant, the material function $\eta$ encodes the dissipative viscous effects and is proportional to the shear or extensional viscosities. 
The interpretation of the material functions $\lambda_0$ and $\lambda_3$ in planar flows is of particular interest (see again Fig.~\ref{fig:material-functions}).
The tensor $\hat{\vt E}$, in the case $\alpha=1/2$, reads
\begin{equation}
\hat{\vt E}
=
-\frac{1}{2}\dv_1\tensprod\dv_1-\frac{1}{2}\dv_2\tensprod\dv_2
+\dv_3\tensprod\dv_3.\label{131021_15Apr18}
\end{equation}
The term $2\ed\lambda_0\hat{\vt E}$ induces only a shift in the eigenvalues of the stress which is isotropic (akin to a pressure) in the flow plane, with a term $-\ed\lambda_0$, while globally anisotropic, since the eigenvalue in the remaining direction is shifted by $2\ed\lambda_0$.
The anisotropy induced in the stress due to this conservative effect can be described by the \emph{ellipsoidal factor} $2\ed\lambda_0/p$.

Meanwhile, the term $2\ed\lambda_3\hat{\vt G}_3$ induces a reorientation of the eigenvectors of the stress tensor with respect to those of $\vt D$ of a \emph{reorientation angle} $\varphi$, such that
\begin{equation}
\tan\varphi=\frac{\lambda_3}{\eta+\sqrt{\eta^2+\lambda_3^2}},
\end{equation}
which is well approximated by the \emph{reorientation factor} $\lambda_3/2\eta$ when $\lambda_3\ll\eta$.
Since, in the planar case, $\hat{\vt E}$ and $\hat{\vt G}_3$ commute, the redistribution of the eigenvalues and the reorientation of the eigenvectors are completely independent effects.

For the case of simple shear flows ($\bar\beta_3=1$), we can easily relate the material functions to the familiar viscometric functions defined for a simple shear with rate $\dot{\gamma}=2\ed$. 
Indeed, the shear viscosity $\eta_\mathrm{S}$ and the normal stress differences $N_1$ and $N_2$ are given by 
\begin{gather}
  \eta_{\mathrm{S}}(\dot{\gamma}=2\ed) = \eta( \ed, \bar{\beta}_3=1 ),\\[2pt]
  N_1(\dot{\gamma}=2\ed) = -4\ed\lambda_3(\ed,\bar{\beta}_3=1) , \\[2pt]
  N_2(\dot{\gamma}=2\ed) = 
  2\ed\lambda_3(\ed,\bar{\beta}_3=1) - 3 \ed\lambda_0 (\ed,\bar{\beta}_3=1).
\end{gather}
We remark again that our representation for $\cauchy$ helps to distinguish between two effects, namely the reorientation of eigenvectors and redistribution of eigenvalues, that can occur independently but are combined in the definition of $N_2$.

These observations show that our set of material functions provides a natural generalization of the classical viscometric functions, which are recovered as specific slices of the former.
Rheological measurements in extensional flows are also reflected in the description of a specific slice of the general material functions, that is the one obtained by setting $\bar\beta_3=0$.
For instance, the conventional value of the planar extensional viscosity is given by $\eta_\mathrm{E}(\ed)=4\eta(\ed,\bar\beta_3=0)$.
Notice that our framework removes the small discrepancies in the standard choices of reference deformation rates and normalization of the viscosities for simple shear and extensional flows, providing consistent definitions for any local flow type.

\section{Reinterpretation of existing models}\label{sec:constitutive}

Here we show the general compatibility of the rheometric framework introduced above with constitutive models. 
As already mentioned, the construction of models requires, as a first step, the selection of the independent quantities upon which the material functions can depend.
Based on different choices we can identify different classes of models.
In what follows, we first reinterpret classical models and then indicate the connection between more recent models and the general decomposition 
of the stress tensor provided in Sec.~\ref{sec:representation}.

\subsection{Response depending on local rate and flow type}

The simplest class of models can be constructed by assuming that the material functions depend only on the local rate $\ed$ and the flow type.
This assumption may seem quite natural when dealing with homogeneous incompressible fluids and is indeed the basis of the most classical fluid models.
It is easy to argue that, if the velocity gradient ought to be the only relevant descriptor, any characteristic relaxation time associated with the microscopic physics of the fluid must be short compared to the time needed to change the local flow type.
This indicates that such an assumption will be effective when the response of the fluid can be practically regarded as instantaneous.
Within this framework, the classical model of Newtonian fluids is obviously recovered by setting $\eta$ constant, independent of any kinematical parameter, and letting all the $\lambda_k$ (for $k=0,\ldots,4$) vanish identically.

\subsubsection{Reiner--Rivlin fluids}

A historically important class of objective models rests on the assumption that the stress depends only on $\vt D$.
It means that the material functions are determined by their values as $\ed$ and $\alpha$ are varied while keeping $\beta_1=\beta_2=\beta_3=0$.
In other words, the fluid behavior is completely characterized by its behavior in purely extensional flows in which the eigenvectors of the stress tensor remain aligned to those of $\vt D$, entailing a uniformly vanishing value of $\lambda_1$, $\lambda_2$, and $\lambda_3$. 

The flow-type dependence in such models is severely restricted and the general expression of the stress becomes
\begin{equation}
\cauchy=-p\vt I+2\eta(\ed,\alpha)\vt D+2\lambda_0(\ed,\alpha)\vt E.
\end{equation}
It is easy to check that these models correspond to the class of Reiner-Rivlin fluids \cite{Macosko_1994,Larson_1999}, for which the stress tensor takes the form
\begin{equation}
\cauchy=-p\vt I+f_1(\mathrm{II,III})\vt D+f_2(\mathrm{II,III})[\vt D^2-(\mathrm{II}/3)\vt I],
\end{equation}
where $f_1$ and $f_2$ are arbitrary scalar functions of the invariants $\mathrm{II}=\tr(\vt D^2)$ and $\mathrm{III}=\det\vt D$.
Indeed, $\mathrm{II}$ and $\mathrm{III}$ can be expressed in terms of $\ed$ and $\alpha$, while $\vt D^2$, being obviously diagonal on the basis of the eigenvectors of $\vt D$, can be written as a linear combination of $\vt I$, $\vt D$, and $\vt E$ with coefficients that depend only on $\ed$ and $\alpha$. 
More explicitly, by applying the definition \eqref{eq:rf} we find
\begin{equation}
2\eta=f_1+f_2\frac{\tr(\vt D^3)}{\mathrm{II}}\quad\text{and}\quad
2\lambda_0=f_2\frac{\vt D^2:\vt E}{\vt E:\vt E}.
\end{equation}

\subsubsection{Models with flow-type dependence}

Since for many fluids, even restricting attention to planar flows, the viscous response in simple shear differs from that in extensional flows, several models have been developed to include a dependence on the flow type.
Here we discuss the connection between our scheme and a few relevant models \cite{Schunk_1990,Souza-Mendes_1995,Hartkamp_2013}.

In the paper of Schunk and Scriven \cite{Schunk_1990} the set of kinematical parameters on which the stress tensor can depend is expanded to include (some of) the degrees of freedom associated with the rate of change of the tensor $\vt D$ along streamlines.
In particular, the local spin of the eigenvectors of $\vt D$ is encoded in the vector $\vc w$ and the parameters $\delta_k$ (for $k=1,2,3$) defined in \eqref{eq:Dspin} and \eqref{eq:dspin}.
The dependence on the flow type is then included in the model by essentially prescribing the material function $\eta$ in terms of $\ed$ and the normalized relative rotation rate encoded in the differences $\bar\beta_k=\beta_k-\delta_k$ (for $k=1,2$, and $3$).

Taking a more general perspective, Souza Mendes~\textit{et. al.}~\cite{Souza-Mendes_1995} consider the symmetric tensor $\vt R=\bar{\vt W}^2$, where the relative rate of rotation tensor is given by
\begin{multline}
\bar{\vt W}
=\ed \left[
(\delta_1-\beta_1)\bigl(\dv_3\tensprod\dv_2-\dv_2\tensprod\dv_3\bigr) 
\right.  \\ \left. 
+(\delta_2-\beta_2)\bigl(\dv_1\tensprod\dv_3-\dv_3\tensprod\dv_1\bigr)
+(\delta_3-\beta_3)\bigl(\dv_2\tensprod\dv_1-\dv_1\tensprod\dv_2\bigr)
\right],
\end{multline}
and then present a general representation of the stress tensor in terms of $\vt D$ and $\vt R$.
Due to the definition of $\vt R$, it is clear that their general representation can be recast in terms of a generic prescription of the material functions $\eta$ and $\lambda_k$ (for $k=0,\ldots,3$) in terms of $\ed$, $\alpha$, and $\bar\beta_k$ (for $k=1,2$, and $3$).

Nevertheless, the particular choice of the form of $\vt R$ imposes additional constraints, with the main implication of a vanishing $\lambda_3$ in simple shear flows, namely the vanishing of first normal stress differences.
This can be easily understood by checking the form of $\vt R$ in a simple shear flow.
Since we are in the presence of a steady flow with uniform gradient, $\bar{\vt W}$ equals $-\vt W$ and we have
\begin{equation}
\vt R=\vt W^2=-\ed^2\beta_3^2\bigl(\dv_1\tensprod\dv_1+\dv_2\tensprod\dv_2\bigr).
\end{equation}
Since $\vt R$ is indeed diagonal on the eigenvectors of $\vt D$, it can be represented on the basis of $\vt I$, $\vt D$, and $\vt E$.
The same happens whenever two of the differences $\beta_k-\delta_k$ vanish.
The stress tensor acquires then a form akin to that for Reiner--Rivlin fluids, but with the important addition of a dependence on $\bar\beta_3$ in the material functions $\eta$ and $\lambda_0$.

Another framework that can be easily recast within our scheme 
is that of Hartkamp \textit{et.\,al.}~\cite{Hartkamp_2013}, originally developed for planar flows.
They discuss general constitutive prescriptions for the pressure tensor $\vt P=-\cauchy$ in terms of a generalized viscosity, which is exactly the material function $\eta$, the lagging angle $\varDelta\phi$ between the eigenvectors of $\vt D$ and those of $\vt P$ in the flow plane, which corresponds to our reorientation angle $\varphi$, and a measure $a$ of the out-of-flow-plane anisotropy of the pressure tensor, that is proportional to our material function $\lambda_0$.
The effectiveness of their framework is tested by building a constitutive model able to nicely capture numerical results for the pressure tensor of a Weeks--Chandler--Andersen fluid in any mixed planar flow.
The importance of the results of Hartkamp~\textit{et.\,al.}~\cite{Hartkamp_2013} should be emphasized by their reinterpretation within the framework introduced in the present paper.
We offer a possibly more flexible and general scheme, but some important ideas are clearly present in their work.

\subsubsection{{Second-order fluids}}

{In the classical models of second-order fluids the stress tensor $\cauchy$ is represented in terms of the first and second Rivlin--Ericksen tensors, respectively $2\vt D$ and $2(\dot{\vt D}+\vt W\cdot\vt D-\vt D\cdot\vt W)$, according to 
\begin{equation}
\cauchy=-p\vt I+2\eta_0\vt D+4\alpha_1\vt D^2+2\alpha_2(\dot{\vt D}+\vt W\cdot\vt D-\vt D\cdot\vt W).\label{130828_15Apr18}
\end{equation}
The term proportional to $\dot{\vt D}$ can have components on any of the elements of the tensorial basis \eqref{eq:basis}.
On the other hand, if we restrict attention to steady homogeneous planar flows ($\dot{\vt D}=\vt 0$), second-order fluids correspond to choosing material functions of the form
\begin{equation}
\eta(\ed,\bar{\beta}_3)=\eta_0,~
\lambda_0(\ed,\bar{\beta}_3)=-\frac{4}{3}\alpha_1\ed,
~\text{and}~\lambda_3(\ed,\bar{\beta}_3)=\alpha_2\ed\bar{\beta}_3,
\end{equation}
since the term $\vt W\cdot\vt D-\vt D\cdot\vt W$ has components only along $\hat{\vt G}_3$.}

\subsection{Response depending on other evolving fields}

The usefulness of the interpretative scheme introduced in Sec.~\ref{sec:representation} goes beyond the construction of models in which the material response depends only on the velocity gradient.
To exemplify this fact we can analyze the models of particulate suspensions proposed 
by Stickel~\textit{et.\,al.}~\cite{Stickel_2006} and Miller~\textit{et.\,al.}~\cite{Miller_2009}.

In \cite{Stickel_2006}, the microstructural properties of the suspension are encoded in a symmetric tensor $\vt Y$ and the effective stress in the fluid is an isotropic polynomial function of $\vt D$ and $\vt Y$. 
The general representation of such a function given in their Eq.~\eqref{130828_15Apr18} can be readily replaced by the following procedure.
First, we represent the six degrees of freedom of the symmetric tensor $\vt Y$ using its three eigenvalues ($y_1$, $y_2$, $y_3$) and the three Euler angles ($\theta_1$, $\theta_2$, $\theta_3$) that identify its eigenvectors with respect to the eigenvectors of $\vt D$.
Then, the general representation of the stress tensor becomes Eq.\,\eqref{eq:T} with all the material functions depending on the set of parameters
\begin{equation}
\mathcal P=\{\ed,\alpha,y_1,y_2,y_3,\theta_1, \theta_2,\theta_3\}.
\end{equation} 
Even though dealing with five arbitrary functions of the parameter set $\mathcal P$ can still be very complicated, the interpretation of the material functions gives a better idea of the role of each term in the stress tensor.
Moreover, any evolution of the microstructure $\vt Y$ can be described independently and then translated into the updated values of the relevant degrees of freedom.

In the paper by Miller~\textit{et.\,al.}~\cite{Miller_2009}, the parameter $|\vc w|$ associated with the relative rotation rate is used to identify the flow type.
The volume fraction $\phi$ of particles in the fluid is another field evolving in the system.
The contribution to the stress due to the presence of the particles is modeled through a dependence of $\eta$ on $\phi$ and a term proportional to the tensor parameter $\vt Q_{ct}$, which is said to represent normal stress differences, with a coefficient that depends on both $\phi$ and $|\vc w|$.
The tensor $\vt Q_{ct}$ given in their Eq.\,\eqref{131021_15Apr18} can be easily seen to be a linear combination of $\vt I$, $\hat{\vt D}$, $\hat{\vt E}$, and $\hat{\vt G}_3$ with coefficients that depend on $\ed$, $\phi$, and $|\vc w|$, since their ``tension-compression coordinates'' are determined exactly by the eigenvectors of $\vt D$.
We thus see how the constitutive model discussed in \cite{Miller_2009} can be used to exemplify the effectiveness of our scheme also in the presence of additional evolving fields such as the volume fraction $\phi$.

\section{Conclusions}

In this article we have introduced a tensorial basis adapted to local flow conditions which can be used to organize data regarding the material response of incompressible fluids in any flow. 
Such a basis is determined, for each point in space and instant in time, by the symmetric part of the velocity gradient.
Within this framework, a description of the effects associated with the independent degrees of freedom of the stress can be easily given.
This supports a coherent interpretation of rheological measurements and computational results obtained under different flow conditions.

The material functions associated with the decomposition of the stress on the adapted tensorial basis generalize and complete the classical set of viscometric functions, which describe the response only in viscometric flows.
The enhanced characterization of the fluid behavior in steady flows can then be used to extrapolate constitutive models for complex fluids starting from rheological data in both viscometric and non-viscometric flows.

Consider, for instance, the contraction flow of Fig.~\ref{fig:contraction}
and assume that we are dealing with a fluid the behavior of which is consistent with the simple flow classification given by $\ed$ and $\bar\beta_3$.
It means that these are the only variables that influence the stress.
We choose such a restricted situation for simplicity, but similar arguments can be extended to more complex fluids.
However, this assumption is reasonable for steady flows such that the timescale over which the flow condition experienced by a fluid parcel changes is long compared to the characteristic time for the stress to reach a steady value in a homogeneous flow.  

In this case, we can experimentally or computationally determine the material functions $\eta$, $\lambda_0$, and $\lambda_3$ by varying $\ed$ and $\bar\beta_3$ in homogeneous steady flows and then use their values to predict the flow everywhere in the contraction geometry.
This would be impossible by using viscometric functions as they can predict the behavior only where the flow is equivalent to simple shear. 
In contrast, the material functions $\eta$, $\lambda_0$, and $\lambda_3$ provide a coherent description of the flow response also in the regions where the local flow condition is of extensional or mixed type.

\section*{Acknowledgments}
The authors acknowledge the support from the Okinawa Institute of Science and Technology Graduate University.
The work of R.~S. was supported by JSPS KAKENHI Grant Number JP17K05618.
The authors also thank the anonymous Reviewers alongside F.\ Del Giudice, M.\ Denn, S.\ Haward, S.\ Janssens, J.\ Morris, H.\ Notsu, and A.\ Shen for providing useful comments on the manuscript.



\begin{thebibliography}{10}

\bibitem{Coleman_1966}
B.~D. Coleman, H.~Markovitz, and W.~Noll.
\newblock {\em Viscometric {F}lows of Non-Newtonian Fluids: Theory and
  Experiment}.
\newblock Springer Verlag, Berlin, 1966.

\bibitem{Macosko_1994}
C.~W. Macosko.
\newblock {\em Rheology}.
\newblock Wiley-VCH, New York, 1994.

\bibitem{Larson_1999}
R.~G. Larson.
\newblock {\em The Structure and Rheology of Complex Fluids}.
\newblock Oxford University Press, New York \& Oxford, 1999.

\bibitem{McKinley_2002}
G.~H. McKinley and T.~Sridhar.
\newblock Filament-stretching rheometry of complex fluids.
\newblock {\em Annu. Rev. Fluid Mech.}, 34(1):375--415, 2002.

\bibitem{Petrie_2006}
C.~J.~S. Petrie.
\newblock One hundred years of extensional flow.
\newblock {\em J. Non-Newtonian Fluid Mech.}, 137(1--3):1--14, 2006.


\bibitem{Dai_2017}
S.~Dai and R.~I. Tanner.
\newblock Elongational flows of some non-colloidal suspensions.
\newblock {\em Rheol. Acta}, 56(1):63--71, 2017.


\bibitem{Kraynik_1992}
A.~M. Kraynik and D.~A. Reinelt.
\newblock Extensional motions of spatially periodic lattices.
\newblock {\em Int. J. Multiphase Flow}, 18(6):1045--1059, 1992.

\bibitem{Todd_1998}
B.~D. Todd and P.~J. Daivis.
\newblock Nonequilibrium molecular dynamics simulations of planar elongational
  flow with spatially and temporally periodic boundary conditions.
\newblock {\em Phys. Rev. Lett.}, 81:1118--1121, 1998.

\bibitem{Baranyai_1999}
A.~Baranyai and P.~T. Cummings.
\newblock Steady state simulation of planar elongation flow by nonequilibrium
  molecular dynamics.
\newblock {\em J. Chem. Phys.}, 110(1):42--45, 1999.

\bibitem{Hunt_2010}
T.~A. Hunt, S.~Bernardi, and B.~D. Todd.
\newblock A new algorithm for extended nonequilibrium molecular dynamics
  simulations of mixed flow.
\newblock {\em J. Chem. Phys.}, 133(15), 2010.

\bibitem{Dobson_2014}
M.~Dobson.
\newblock Periodic boundary conditions for long-time nonequilibrium molecular
  dynamics simulations of incompressible flows.
\newblock {\em J. Chem. Phys.}, 141(18):184103, 2014.

\bibitem{Zinchenko_2015}
A.~Z. Zinchenko and R.~H. Davis.
\newblock Extensional and shear flows, and general rheology of concentrated
  emulsions of deformable drops.
\newblock {\em J. Fluid Mech.}, 779:197--244, 2015.

\bibitem{Jain_2015}
A.~Jain, C.~Sasmal, R.~Hartkamp, B.~D. Todd, and J.~R. Prakash.
\newblock Brownian dynamics simulations of planar mixed flows of polymer
  solutions at finite concentrations.
\newblock {\em Chem. Eng. Sci.}, 121:245--257, 2015.

\bibitem{Hunt_2016}
T.~A. Hunt.
\newblock Periodic boundary conditions for the simulation of uniaxial
  extensional flow of arbitrary duration.
\newblock {\em Mol. Simulat.}, 42(5):347--352, 2016.

\bibitem{Cheal_2018}
O.~Cheal and C.~Ness.
\newblock Rheology of dense granular suspensions under extensional flow.
\newblock {\em J. Rheol.}, 62(2):501--512, 2018.

\bibitem{taylor_1934}
G.~I. Taylor.
\newblock The formation of emulsions in definable fields of flow.
\newblock {\em Proc. R. Soc. Lond. A Mat.}, 146:501--523, 1934.

\bibitem{Fuller_1981}
G.~G. Fuller and L.~G. Leal.
\newblock The effects of conformation-dependent friction and internal viscosity
  on the dynamics of the nonlinear dumbbell model for a dilute polymer
  solution.
\newblock {\em J. Non-Newtonian Fluid Mech.}, 8(3):271--310, 1981.

\bibitem{Hudson_2004}
S.~D. Hudson, F.~R. Phelan~Jr., M.~D. Handler, J.~T. Cabral, K.~B. Migler, and
  E.~J. Amis.
\newblock Microfluidic analog of the four-roll mill.
\newblock {\em Appl. Phys. Lett.}, 85(2):335--337, 2004.

\bibitem{Lee_2007}
J.~S. Lee, R.~Dylla-Spears, N.~P. Teclemariam, and S.~J. Muller.
\newblock Microfluidic four-roll mill for all flow types.
\newblock {\em Appl. Phys. Lett.}, 90(7):074103, 2007.

\bibitem{Lee_2007a}
J.~S. Lee, E.~S.~G. Shaqfeh, and S.~J. Muller.
\newblock Dynamics of {DNA} tumbling in shear to rotational mixed flows:
  {P}athways and periods.
\newblock {\em Phys. Rev. E}, 75:040802, 2007.

\bibitem{Deschamps_2009}
J.~Deschamps, V.~Kantsler, E.~Segre, and V.~Steinberg.
\newblock Dynamics of a vesicle in general flow.
\newblock {\em Proc. Natl. Acad. Sci. USA}, 106(28):11444--11447, 2009.

\bibitem{Haward_2012}
S.~J. Haward, M.~S.~N. Oliveira, M.~A. Alves, and G.~H. McKinley.
\newblock Optimized cross-slot flow geometry for microfluidic extensional
  rheometry.
\newblock {\em Phys. Rev. Lett.}, 109:128301, 2012.

\bibitem{Fuller_1995}
G.~G. Fuller.
\newblock {\em Optical Rheometry of Complex Fluids}.
\newblock Oxford University Press, New York, 1995.

\bibitem{Pathak_2006}
J.~A. Pathak and S.~D. Hudson.
\newblock Rheo-optics of equilibrium polymer solutions: {W}ormlike micelles in
  elongational flow in a microfluidic cross-slot.
\newblock {\em Macromolecules}, 39(25):8782--8792, 2006.

\bibitem{Ober_2013}
T.~J. Ober, S.~J. Haward, C.~J. Pipe, J.~Soulages, and G.~H. McKinley.
\newblock Microfluidic extensional rheometry using a hyperbolic contraction
  geometry.
\newblock {\em Rheol. Acta}, 52(6):529--546, 2013.

\bibitem{Shribak_2015}
M.~Shribak.
\newblock Polychromatic polarization microscope: bringing colors to a colorless
  world.
\newblock {\em Sci. Rep.}, 5, 2015.

\bibitem{Zhao_2016}
Y.~Zhao, A.~Q. Shen, and S.~J. Haward.
\newblock Flow of wormlike micellar solutions around confined microfluidic
  cylinders.
\newblock {\em Soft Matter}, 12:8666--8681, 2016.

\bibitem{Sun_2016}
C.-L. Sun and H.-Y. Huang.
\newblock Measurements of flow-induced birefringence in microfluidics.
\newblock {\em Biomicrofluidics}, 10(1):011903, 2016.

\bibitem{Callaghan_1999}
P.~T. Callaghan.
\newblock Rheo-{NMR}: nuclear magnetic resonance and the rheology of complex
  fluids.
\newblock {\em Rep. Prog. Phys.}, 62:599--670, 1999.

\bibitem{Besseling_2007}
R.~Besseling, E.~R. Weeks, A.~B. Schofield, and W.~C.~K. Poon.
\newblock Three-dimensional imaging of colloidal glasses under steady shear.
\newblock {\em Phys. Rev. Lett.}, 99:028301, 2007.

\bibitem{Manneville_2008}
S.~Manneville.
\newblock Recent experimental probes of shear banding.
\newblock {\em Rheol. Acta}, 47(3):301--318, 2008.

\bibitem{Dimitriou_2012}
C.~J. Dimitriou, L.~Casanellas, T.~J. Ober, and G.~H. McKinley.
\newblock Rheo-{PIV} of a shear-banding wormlike micellar solution under large
  amplitude oscillatory shear.
\newblock {\em Rheol. Acta}, 51(5):395--411, 2012.

\bibitem{Gallot_2013}
T.~Gallot, C.~Perge, V.~Grenard, M.-A. Fardin, N.~Taberlet, and S.~Manneville.
\newblock Ultrafast ultrasonic imaging coupled to rheometry: Principle and
  illustration.
\newblock {\em Rev. Sci. Instrum.}, 84(4):045107, 2013.

\bibitem{Saint-Michel_2016}
B.~Saint-Michel, T.~Gibaud, M.~Leocmach, and S.~Manneville.
\newblock Local oscillatory rheology from echography.
\newblock {\em Phys. Rev. Applied}, 5:034014, 2016.

\bibitem{Seto_2017}
R. Seto, G.~G. Giusteri, and A. Martiniello.
\newblock Microstructure and thickening of dense suspensions under extensional
  and shear flows.
\newblock {\em J. Fluid Mech.}, 825:R3, 2017.

\bibitem{Pasquali_2004}
M. Pasquali and L.~E. Scriven.
\newblock Theoretical modeling of microstructured liquids: a simple thermodynamic approach.
\newblock {\em J. Non-Newtonian Fluid Mech.}, 120(1--3):101--135, 2004.

\bibitem{Pasquali_2002}
M. Pasquali and L.~E. Scriven.
\newblock Free surface flows of polymer solutions with models based on the conformation tensor.
\newblock {\em J. Non-Newtonian Fluid Mech.}, 108(1--3):363--409, 2002.

\bibitem{Thompson_2005}
R.~L. Thompson and P.~R. Souza~Mendes.
\newblock Considerations on kinematic flow classification criteria.
\newblock {\em J. Non-Newtonian Fluid Mech.}, 128(2--3):109--115, 2005.

\bibitem{Astarita_1979}
G.~Astarita.
\newblock Objective and generally applicable criteria for flow classification.
\newblock {\em J. Non-Newtonian Fluid Mech.}, 6:69--76, 1979.

\bibitem{Beris_1994}
A.~N. Beris and B.~J. Edwards.
\newblock {\em Thermodynamics of Flowing Systems with Internal Microstructure}.
\newblock Oxford University Press, New York \& Oxford, 1994.

\bibitem{Phan-Thien_2002}
N. Phan-Thien.
\newblock {\em Understanding Viscoelasticity. Basics of Rheology}.
\newblock Springer-Verlag, Berlin \& Heidelberg, 2002.

\bibitem{Ryskin_1980}
G. Ryskin and J.~M. Rallison.
\newblock The extensional viscosity of a dilute suspension of spherical particles at intermediate microscale Reynolds numbers.
\newblock {\em J. Fluid Mech.}, 99(3):513--529, 1980.

\bibitem{Schunk_1990}
P.~R. Schunk and L.~E. Scriven.
\newblock Constitutive equation for modeling mixed extension and shear in
  polymer solution processing.
\newblock {\em J. Rheol.}, 34(7):1085--1119, 1990.

\bibitem{Haward_2016}
S.~J. Haward, G.~H. McKinley, and A.~Q. Shen.
\newblock Elastic instabilities in planar elongational flow of monodisperse
  polymer solutions.
\newblock {\em Sci. Rep.}, 6:33029, 2016.

\bibitem{Souza-Mendes_1995}
P.~R. Souza~Mendes, M.~Padmanabhan, L.~E. Scriven, and C.~W. Macosko.
\newblock Inelastic constitutive equations for complex flows.
\newblock {\em Rheol. Acta}, 34(2):209--214, 1995.

\bibitem{Hartkamp_2013}
R.~Hartkamp, B.~D. Todd, and S.~Luding.
\newblock A constitutive framework for the non-{N}ewtonian pressure tensor of a
  simple fluid under planar flows.
\newblock {\em J. Chem. Phys.}, 138(24):244508, 2013.

\bibitem{Stickel_2006}
J.~J. Stickel, R.~J. Phillips, and R.~L. Powell.
\newblock A constitutive model for microstructure and total stress in
  particulate suspensions.
\newblock {\em J. Rheol.}, 50:379--413, 2006.

\bibitem{Miller_2009}
R.~M. Miller, J.~P. Singh, and J.~F. Morris.
\newblock Suspension flow modeling for general geometries.
\newblock {\em Chem. Eng. Sci.}, 64:4597--4610, 2009.

\end{thebibliography}
\end{document}